\documentclass[useAMS,usenatbib,a4paper]{mn2e}

\usepackage{graphicx}
\usepackage{times}
\newcommand{\cthead}[1]{\multicolumn{1}{c}{#1}}
\newcommand{\kss}{km~s$^{-1}$ }
\newcommand{\ks}{km~s$^{-1}$}
\newcommand{\astrophtext}{This is a preprint of an article accepted for
publication in MNRAS \copyright\ 2005 RAS}
\makeatletter
\renewcommand{\@oddfoot}{\small \astrophtext\hfil}
\renewcommand{\@evenfoot}{\hfil\small\astrophtext}
\def\ps@titlepage{\let\@mkboth\@gobbletwo
\def\@oddhead{\footnotesize\@journal\hfill}
\def\@oddfoot{\small \astrophtext\hfil}
\def\@evenhead{\footnotesize\@journal\hfill}
\def\@evenfoot{\hfil\small\astrophtext}
}
\makeatother

\title[The 6.7-GHz and 25-GHz methanol masers in OMC-1]
{The 6.7-GHz and 25-GHz methanol masers in OMC-1}
\author[M. A. Voronkov et al.]{M. A. Voronkov$^{1,2}$\thanks{E-mail:
Maxim.Voronkov@csiro.au}, A. M. Sobolev$^{3}$, S.P. Ellingsen$^{4}$,
A.B.Ostrovskii$^{5}$\\
$^{1}$Australia Telescope National Facility CSIRO, Locked Bag 194, Narrabri,
NSW 2390, Australia\\
$^{2}$Astro Space Centre, Profsouznaya st. 84/32, 117997 Moscow, Russia\\
$^{3}$Astronomical Observatory of the Ural State University, Lenin pr-t. 51,
620083 Ekaterinburg, Russia\\
$^{4}$School of Mathematics and Physics, University of Tasmania, GPO Box
252-37, Hobart, Tasmania 7000, Australia\\
$^{5}$Ural State University, Lenin pr-t. 51, 620083 Ekaterinburg, Russia\\}
\begin{document}

\date{}

\pagerange{\pageref{firstpage}--\pageref{lastpage}} \pubyear{2004}

\maketitle

\label{firstpage}

\begin{abstract}

  The Australia Telescope Compact Array (ATCA) has been used to search
  for methanol maser emission at 6.7~GHz towards OMC-1. Two features
  peaking at 7.2~\kss and $-$1.1~\kss have been detected. The former
  has at least two components close in both velocity and position. It
  is located south-east of the Orion Kleinmann-Low (Orion-KL) nebula
  in the region of outflow traced by the 25-GHz methanol masers and
  the 95-GHz methanol emission. It is shown by modelling that in
  contrast to the widespread opinion that simultaneous masing of
  methanol transitions of different classes is impossible there are
  conditions for which simultaneous masing of the class~II transition
  at 6.7-GHz and some class~I transitions (e.g. the series at 25~GHz)
  is possible. A relevant example is provided, in which the pumping
  occurs via the first torsionally excited state and is driven by
  radiation of the dust intermixed with the gas in the cloud.  In this
  regime the dust temperature is significantly lower (T$\approx$60~K)
  than in the case of bright 6.7-GHz masers (T$>$150~K). The narrow
  spectral feature at $-$1.1~\kss has a brightness temperature greater
  than about 1400~K, which suggests that it is probably a maser. It
  emanates from the Orion South region and is probably associated with
  the approaching part of outflow seen in CO.  The 25-GHz maser
  associated with OMC-1 was observed quasi-simultaneously with the
  6.7-GHz observations. No 25-GHz emission associated with the
  $-$1.1~\kss 6.7~GHz feature towards Orion South was detected.
\end{abstract}


\section{Introduction}
The Orion Molecular Cloud 1 (OMC-1) is a well known region of massive
star formation containing several young stellar objects (YSOs) at the
very early stages of their formation as well as outflows related to
them \citep[see, e.g. the review by][]{gen89}.  Methanol masers in a
variety of transitions are often found in such regions. Masing
transitions are traditionally divided into two classes. Although
masers of both classes often co-exist in the same star-forming region,
they are usually seen apart from each other \citep{men91a,men91b}.

OMC-1 harbours strong class~I methanol masers. These masers are
believed to trace distant parts of the outflows from YSOs and are
associated with the early stages of the massive star formation
\citep[e.g.,][]{men91a,ell05}. The strongest (about 150 Jy) masers of
the J$_2-$J$_1$~E line series near 25 GHz are observed in OMC-1 and
have inspired numerous observational and theoretical studies
\citep[see, e.g.,][]{men88b,joh92,sob03}.

In contrast, strong class II masers have not been detected in OMC-1.
Only weak emission in the strongest class~II $5_1-6_0$~A$^+$
transition at 6.7~GHz has been reported from this source by
\citet{cas95}, and it was unclear whether the observed spectral line
is a maser or has a quasi-thermal origin (i.e., the level populations
are without inversion). Currently, class~II methanol masers are
thought to appear in the vicinity of the massive young stellar objects
(YSOs) at the early stages of ultra-compact {\sc HII} region (UCHII)
evolution \citep[e.g.,][]{ell96,min01,cod04}.  The transition at
6.7~GHz manifests flux densities which are the highest among methanol
masers and can reach 5000~Jy. Maser emission from this transition is
widespread and to date has been detected towards more than 500 sources
\citep[see, e.g. the compilation by][]{mal03}. Being the strongest
this transition is the best prospect for searches for new class II
methanol maser sites.

Evolutionary stages of the massive YSOs at which different classes of
methanol masers exist do intersect considerably. \citet{ell05}
detected class~I methanol masers at 95.1~GHz towards 38\% of a
statistically complete sample of the class~II masers at 6.7~GHz.  No
extensive search of class~II masers has been performed towards the
positions of isolated class~I masers. This is partly due to the lack
of information about such sources since most class~I masers have been
found towards known class~II masers, which are often located within a
single dish beam \citep[e.g.,][]{sly94}.

Therefore, among known sources OMC-1 is somewhat exceptional: it
posesses strong class~I masers produced by an outflow \citep{joh92},
while there are no strong class~II masers associated with the YSO
producing this outflow. Recent observations have shown that some
hydroxyl masers and class~II methanol masers can also be associated
with outflows \citep{arg03,cod04}. This is supported by statistical
analysis of correlation between the maser velocities and those of the
molecular outflow tracers \citep{mal03} and is reasonable because the
methanol abundance is significantly increased in the interface
regions, where the outflow interacts with the ambient material
\citep[e.g.,][]{bac95,bac98,gib98}.  It is presently unclear whether
weak class~II masers can co-exist with the class~I masers or they
always trace different parts of outflow.  We have undertaken a
sensitive search at 6.7~GHz in order to figure out whether there are
weak class~II masers in the vicinity of the YSOs and outflows in
OMC-1. The 25-GHz class~I maser in OMC-1 was reobserved to check its
current status.

\section{Observations}
\begin{table}
\caption{Dates of observations and a summary of array configurations.
The compact H75B configuration featured the north-south
spur of the ATCA providing a better uv-coverage for equatorial
sources.  In the 1.5D configuration session the {\sc CA05} antenna did
not observe and, therefore, the second shortest baseline {\sc
CA04--CA05} was not available. Other sessions had all 6 antennae
working.}
\label{obsparam}
\begin{tabular}{@{}l@{ }rcrrr}
\hline
\multicolumn{2}{c}{Date}&Array&\multicolumn{3}{c}{Baseline length (m)}\\
\multicolumn{2}{c}{UT}&configuration&\cthead{Min}&\cthead{Max}&\cthead{Max}\\
\multicolumn{4}{c}{}&\cthead{no CA06}&\cthead{with CA06}\\
\hline
9$-$10 June&2003 & H75B  &31 & 89 & 4408\\
12 November&2003 & 1.5D &107 & 1439 & 4439\\
5 January &2004 & 6A &337 & 2923& 5939 \\
10 April &2004 & EW367 &46 & 367 & 4408 \\
\hline
\end{tabular}
\end{table}

The observations were made using the Australia Telescope Compact Array
(ATCA). Details on the array configurations\footnote{More
details on the ATCA configurations and baseline length distribution
are available on the web {\it
http://www.narrabri.atnf.csiro.au/operations/\\array\_configurations/configurations.html}}
and observing dates are summarized in Table~\ref{obsparam}.
During the first session on 2003 June 9--10, the 25-GHz maser was
observed quasi-simultaneously with the 6.7-GHz observations in 7 short
scans switching the frequency approximately every 90 minutes. The
primary aim of the subsequent sessions was to estimate the 6.7-GHz
source size and therefore the 25-GHz transition was not observed. The
adopted rest frequencies are 24959.080~MHz and 6668.5192~MHz for
$5_2-5_1$~E and $5_1-6_0$~A$^+$ methanol transitions respectively
\citep{lov92,bre95}.  For both frequencies and all sessions the
correlator was configured to split the 4-MHz bandwidth into 1024
spectral channels yielding a spectral resolution of 3.9~kHz or
0.05~\kss at 25~GHz and 0.18~\kss at 6.7~GHz (no Hanning smoothing has
been applied). Two orthogonal linear
polarizations were observed at each frequency, except the {\sc CA01}
antenna in the first session on 2003 June 9--10 which observed only
horizontal polarization (at both frequencies) due to technical
problems. During the data reduction the source has been assumed to be
unpolarized and both polarizations, if presented, were averaged
together.  The position of the phase and pointing centre was
$\alpha_{2000}=5^h35^m$15\fs069,
$\delta_{2000}=-5$\degr23\arcmin42\farcs53 for the first session and
$\alpha_{2000}=5^h35^m$15\fs439,
$\delta_{2000}=-5$\degr23\arcmin07\farcs0 for the latter observations.
The full width at half maximum (FWHM) of the primary beam was 7\farcm2
at 6.7~GHz and 2\farcm0 at 25~GHz. The global pointing solution was
used for both frequencies giving an rms pointing accuracy of
5$-$10\arcsec.

\begin{figure*}
\includegraphics[width=\linewidth]{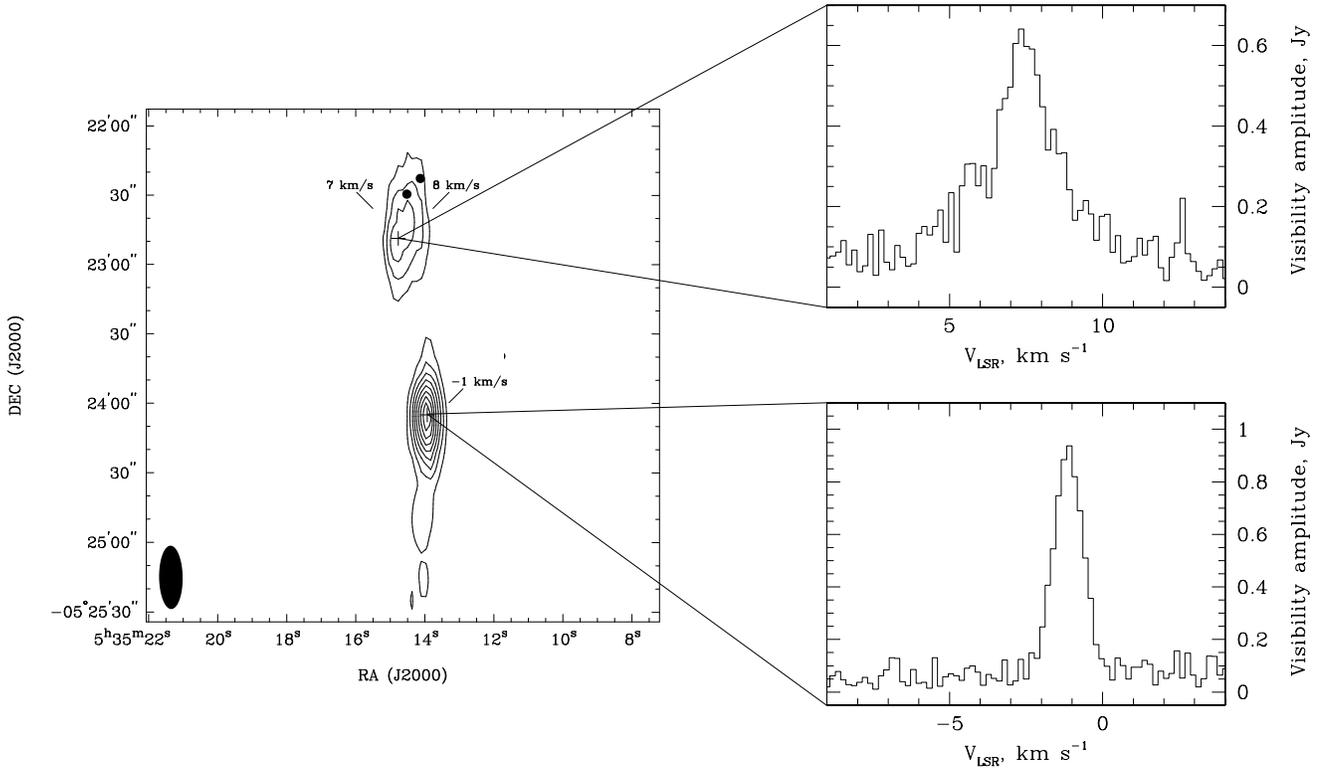}
\caption{A cumulative image of the 6.7-GHz methanol emission in OMC-1
  (all data except those resulted from the {\sc CA06} antenna have been used)
  and the spectra (a vector averaging of all data corresponding to uv-distances less than
  5~k$\lambda$ has been used) at the peak positions for each feature. The filled ellipse
  represent the FWHM of the synthesized beam of
  27\farcs4$\times$10\farcs2 and its position angle of 0\fdg8.
  Contours are 20, 30, 40, 50, 60, 70, 80 and 90 per cent of
  0.49~Jy~beam$^{-1}$. The two filled circles represent IRc2 (lower) and BN
  (upper) objects.}
\label{orion6specandmap}
\end{figure*}
\begin{table*}
\caption{Results of the Gaussian fitting to 6.7-GHz images of individual
spectral channels. All uncertainties are formal errors of the Gaussian fit.
A good estimate of the actual 1$\sigma$ positional accuracy is
5\arcsec$\times$2\arcsec{ }(full width) with a positional angle of
0\fdg8 (the synthesized beam has the FWHM of 27.4\arcsec$\times$10.2\arcsec{ }and
the same positional angle).}
\label{fit6}
\begin{tabular}{@{}rlllrllllll}
\hline
\cthead{LSR}&\cthead{$\alpha_{2000}$}&\cthead{$\delta_{2000}$}&
\cthead{Peak}&\cthead{Total}&\multicolumn{3}{c}{Fitted Gaussian}&
\multicolumn{3}{c}{Deconvolved Gaussian}\\
\cthead{Velocity}&\cthead{5$^h$35$^m$}
&\cthead{$-$5\degr}&\cthead{flux}&\cthead{flux}&\cthead{maj}&\cthead{min}&\cthead{PA}&
\cthead{maj}&\cthead{min}&\cthead{PA}\\
\cthead{(\ks)}&\cthead{($^s$)}&\cthead{(\arcmin~\arcsec)}&
\cthead{(Jy)}&\cthead{(Jy)}&\cthead{(\arcsec)}&\cthead{(\arcsec)}&\cthead{(\degr)}&
\cthead{(\arcsec)}&\cthead{(\arcsec)}&\cthead{(\degr)}\\
\hline
$-$1.1 & 13.93$\pm$0.02 & 24\arcmin05\farcs3$\pm$0\farcs9 & 0.45$\pm$0.03 & 0.52 &
 31$\pm$3 & 10$\pm$1 & \hphantom{$-$}1$\pm$3 & \multicolumn{3}{c}{unresolved}\\
7.2 & 14.77$\pm$0.11& 22\arcmin49\arcsec$\pm$6 & 0.20$\pm$0.03 & 0.54 &
 53$\pm$20 & 14$\pm$6 & $-$7$\pm$9 & 45$\pm$23 & 9$\pm$10 & $-$9$\pm$24 \\
7.3 & 14.60$\pm$0.05& 22\arcmin37\arcsec$\pm$2 & 0.19$\pm$0.02 & 0.48 &
 44$\pm$7 & 16$\pm$3 & $-$13$\pm$6 & 35$\pm$9 & 10$\pm$5 & $-$21$\pm$19\\
7.5 & 14.53$\pm$0.05& 22\arcmin42\arcsec$\pm$2 & 0.22$\pm$0.02 & 0.57 &
 43$\pm$8 & 17$\pm$3 & \hphantom{$-$}0$\pm$6 & 33$\pm$10 & 14$\pm$4 & $-$1$\pm$21\\
7.9 & 14.37$\pm$0.06 & 22\arcmin46\arcsec$\pm$2 & 0.17$\pm$0.01 & 0.42 &
 40$\pm$8 & 17$\pm$4 & $-$5$\pm$8 & 29$\pm$11 & 13$\pm$5 & $-$10$\pm$32 \\
\hline
\end{tabular}
\end{table*}

Most of the data reduction was performed using the {\sc miriad}
package (13 May 2004 release) following standard procedures with the
exception of bandpass calibration for the 6.7-GHz observations. We
found that the accuracy of the bandpass calibration limited our
ability to detect a weak spectral line.  The shape of the uncalibrated
bandpass was reasonably straight within the small part of the band
occupied by the source spectrum and it was therefore safe to skip the
bandpass calibration. However, this approach fails on the shortest
baselines. The continuum emission is very strong towards OMC-1 at
short spacings. Therefore, it is difficult to subtract this continuum
accurately without a proper bandpass calibration, which would require
unrealistically long observations of the bandpass calibrator. A
residual error in the continuum subtraction tends to create weak and
broad spurious lines in the spectrum. A similar problem makes it
extremely difficult to observe a weak spectral line towards a strong
continuum source using a single dish. Therefore, all the 6.7-GHz data
resulting from baselines with a projected spacing shorter than 50m
were flagged. This has a serious impact on the 6.7-GHz observations
using the H75B array configuration, where about one half of the data
was discarded.  Shadowing was also a problem in the compact H75B array
configuration throughout the whole session. All the 25-GHz data
from baselines with a shadowed antenna have been flagged
(3 per cent of visibilities from the total number of baselines without
the {\sc CA06} antenna). The rejection of short spacings at 6.7~GHz
discards all shadowed baselines automatically. During imaging, natural
weighting was used to minimize the noise in the image and to provide
better stability in the deconvolution algorithm ({\sc clean}) for the
near-equatorial source.  The data resulting from the long baselines
with the CA06 antenna were used for non-imaging analysis only.  In all
sessions except the first one the standard ATCA primary calibrator
{\sc 1934-638} was used for the amplitude calibration.  For the first
session {\sc 0605-085} with the adopted flux density of 1.8~Jy at
6.7~GHz was used instead due to the siderial time range allocated.
The 25-GHz data were corrected for the atmospheric opacity using
archival meteorological data and the flux density scale was calibrated
using observations of Uranus.  The {\sc plboot} task of {\sc miriad}
was used to compute the scale factor from the planetary observations
using a built-in ephemeris. For the date of 25-GHz observations it
implied the total flux density of Uranus equal to 0.76~Jy. The
continuum source {\sc 0539-057} was used as a secondary calibrator at
both frequencies.  Although the calibration and imaging were performed
in the {\sc miriad}, the {\sc aips++} package developed by
NRAO\footnote{The National Radio Astronomy Observatory (NRAO) is a
  facility of the National Science Foundation operated under
  cooperative agreement by Associated Universities, Inc.} was used for
a direct manipulation of the 6.7-GHz data to investigate the behavior
of the visibility amplitude at different uv-distances.

\section{Results}
\subsection{The 6.7-GHz data}
\label{result6}

To analyze the spatial and velocity distribution of the 6.7-GHz
emission in OMC-1 we constructed a spectral cube and vector averaged
spectra at different positions. All data from all sessions, except
those including the {\sc CA06} antenna were used to produce the cube.
A vector averaging (i.e. averaging of complex visibilities with phase)
was used for the spectra because the signal was not strong enough for
other methods of presentation. To avoid decorrelation in the spectra
only data corresponding to uv-distances less than 5~k$\lambda$ were
included. A cumulative image of the source obtained from the
cube is shown in Fig.~\ref{orion6specandmap} along with the spectra.
The brightness of each pixel is the maximum across the velocity axis
(different planes of the cube correspond to different spectral
channels and, thus, to different velocities). The image in
Fig.~\ref{orion6specandmap} shows that there are two sources separated
by about 80\arcsec{ }in the north-south direction. Given the positions
of the sources, the reduction in amplitude due to primary beam
attenuation is estimated to be less than 7 per cent, which is less
than the precision of the absolute flux density calibration.  This
effect can therefore be neglected at 6.7~GHz and no primary beam
correction has been applied to either the spectra or the image.  These
sources in the map in Fig.~\ref{orion6specandmap} correspond to two
spectral features, a broad one peaking at 7.2~\kss and a narrow one
peaking at $-$1.1~\ks. An inspection of individual planes of the
spectral cube confirms that no emission brighter than
0.15~Jy~beam$^{-1}$ is visible at the position of one feature within
the velocity range of the other.  Both features were detected in the
vector averaged spectra constructed separately for each individual
observing session, which can be considered as an independent
confirmation.  Radial velocities in the range 4$-$9~\kss are typical
for the Orion region.  The emission from the hot core component peaks
in the range 3$-$8~\kss, while the methanol emission core (compact
ridge) peaks around 7$-$9~\kss\citep{men86a,wri96,cra01}.
\citet{men88a} found the methanol emission in the southern source
peaking at approximately 6~\ks.  However, no methanol emission has been
previously reported at or near $-$1.1~\ks.

The northern source (the feature peaking at 7.2~\ks) is located about
10\arcsec{ } to the south of the IRc2 object. This position is typical
for methanol lines at these velocities and is known as the compact
ridge~\citep{men88a,men88b,wri96}. From Fig.~\ref{orion6specandmap} the
source appears extended and two components are readily distinguished.
The spectrum also supports this as the profile of the broad feature is
asymmetric.  To reveal the velocity structure of the source we fitted
a Gaussian source model to individual planes of the cube used to
construct a cumulative map shown in Fig.~\ref{orion6specandmap}. The
results of these fits are listed in Table~\ref{fit6}. The first five
columns are the LSR velocity of the channel, the position and the peak
flux density of the fitted Gaussian and the total flux density of the
source.  The last six columns are the Gaussian parameters, the length
of the major and minor axes (at half maximum) and the position angle
of the major axis, for the fitted Gaussian and the result of
deconvolution of the synthesized beam. The uncertainties listed in the
table are the formal errors from the Gaussuan fit. The formal errors
for the deconvolved Gaussians are obtained under assumption that the
synthesized beam is known precisely. A good estimate of the actual
1$\sigma$ positional accuracy is the synthesized beam size divided by
the signal to noise ratio in the image \citep[e.g.,][]{fom99}.  The
noise level in the map of each spectral channel (cumulative map is
shown in Fig.~\ref{orion6specandmap}) is about 20 per cent of maximum.
The FWHM of the synthesized beam is 27\farcs4$\times$10\farcs2 and the
position angle is 0\fdg8, which gives an estimate of the 1$\sigma$
positional accuracy of 5\arcsec$\times$2\arcsec{ } with the same
position angle.  Note, however, that for a blended component like
7$-$8~\kss feature is, the actual accuracy may be worse. The fitted
positions in the Table~\ref{fit6} show a trend of moving towards the
east as the velocity decreases from 8~\kss to 7~\ks.
This supports the idea that the source is a blend of two components
with similar velocities and positions. The southern source
($-$1.1~\kss feature, Fig.~\ref{orion6specandmap}) is located at the
position of the Orion-South region. This region is a known source of
methanol emission \citep{men88a}, however, the thermal emission peaks
at a velocity of approximately 6~\ks. The results of the Gaussian fit
into the images of different spectral channels across the line profile
of the $-$1.1~\kss feature are the same within the fitting
uncertainty.  Therefore, in the Table~\ref{fit6} we listed only one of
the fits, corresponding to the peak velocity.
\begin{figure}
\includegraphics[width=\linewidth]{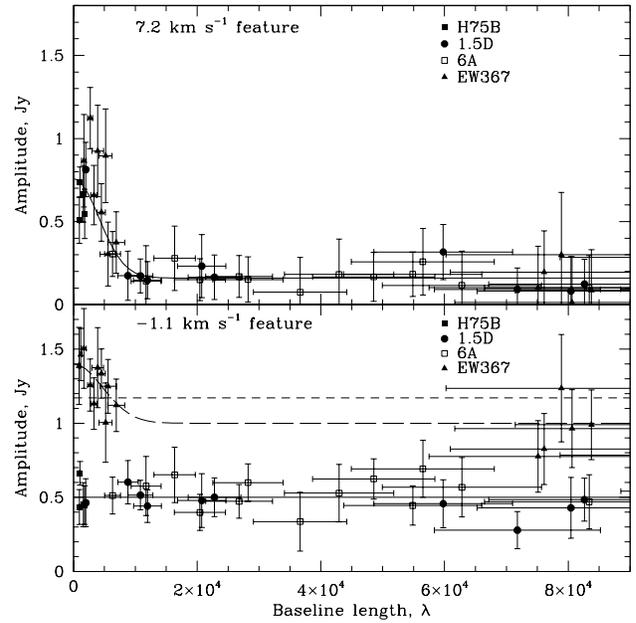}
\caption{The dependence of the visibility amplitude on the baseline
length for 7.2~\kss (upper plot) and $-$1.1~\kss (lower plot) 6.7-GHz features.
Different symbols correspond to the data from different array configurations.
The solid and dashed lines represent the results of model fitting.
The EW367 data for the $-$1.1~\kss feature were fitted
separately (dashed lines) from the data from other configurations (solid line).}
\label{uvplot}
\end{figure}

\begin{figure*}
\includegraphics[width=\linewidth]{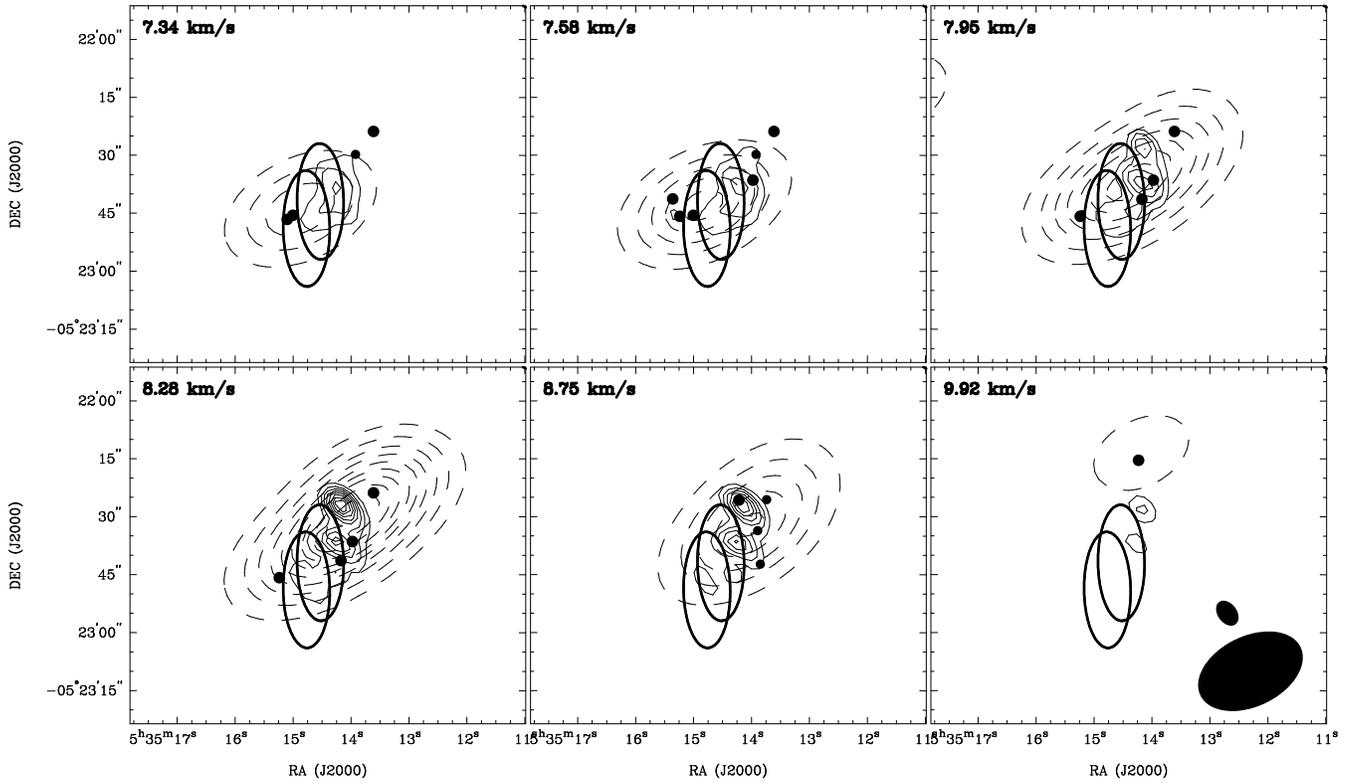}
\caption{Images of the 25-GHz maser (dashed contours) in OMC-1 at different
velocities overlayed on the images of the 95-GHz emission (solid contours)
obtained by \protect\citet{wri96}, courtesy of ADIL. The filled ellipses in the
bottom right corner represent the synthesized beams at 25~GHz
(large ellipse, 28\farcs7$\times$18\farcs1, $-$64\degr) and 95~GHz (small
ellipse, 7.1$\times$4.9, 36\degr).
Contours are 10, 20, 30, 40, 50, 60, 70, 80 and 90 per cent of 86.8~Jy~beam$^{-1}$
at 25~GHz and 20, 30, 40, 50, 60, 70, 80 and 90 per cent of 40.4~Jy~beam$^{-1}$
at 95~GHz. The spectral resolution is 0.5~\kss and 0.05~\kss for
the 95-GHz data obtained by \protect\cite{wri96} and our 25-GHz data,
respectively. Bold open ellipses represent the 3$\sigma$ uncertainty
of the 6.7-GHz emission position and correspond to the fit results
at 7.2~\kss and 7.5~\kss for the left and right ellipse respectively.
The filled
circles represent 25-GHz maser spots revealed by \citet{joh92}. For each
subplot, only spots with the peak velocity closer to the subplot velocity than
their half width at half maximum plus the width of one VLA spectral channel
(0.15~\ks) are displayed. The circles of a smaller radius correspond to
faint 25-GHz spots (less than 10~Jy of the total flux).}
\label{orion25maps}
\end{figure*}

\begin{table*}
\caption{Results of Gaussian fits to 25-GHz images of characteristic
channels shown in Fig.\protect\ref{orion25maps}. All uncertainties are
formal errors of the Gaussian fit. A good estimate of the actual 1$\sigma$
positional accuracy is 2.4\arcsec$\times$1.5\arcsec{ }(full width) with a
positional angle of $-$64\degr (the synthesized beam has the
FWHM of 28.7\arcsec$\times$18.1\arcsec{ }and the same positional angle).}
\label{fit25}
\begin{tabular}{@{}rlllrllllll}
\hline
\cthead{LSR}&\cthead{$\alpha_{2000}$}&\cthead{$\delta_{2000}$}&
\cthead{Peak}&\cthead{Total}&\multicolumn{3}{c}{Fitted Gaussian}&
\multicolumn{3}{c}{Deconvolved Gaussian}\\
\cthead{Velocity}&\cthead{5$^h$35$^m$}
&\cthead{$-$5\degr22\arcmin}&\cthead{flux}&\cthead{flux}&\cthead{maj}&\cthead{min}&\cthead{PA}&
\cthead{maj}&\cthead{min}&\cthead{PA}\\
\cthead{(\ks)}&\cthead{($^s$)}&\cthead{(\arcsec)}&
\cthead{(Jy)}&\cthead{(Jy)}&\cthead{(\arcsec)}&\cthead{(\arcsec)}&\cthead{(\degr)}&
\cthead{(\arcsec)}&\cthead{(\arcsec)}&\cthead{(\degr)}\\
\hline
7.34 & 14.87$\pm$0.03& 44.0$\pm$0.4 & 35$\pm$2 & 37 &
 29.6$\pm$1.1 & 18.6$\pm$0.7 & $-$63$\pm$3 & 7$\pm$4 & 4$\pm$3 &$-$50$\pm$87\\
7.58 & 14.83$\pm$0.03& 42.8$\pm$0.4 & 48$\pm$2 & 51 &
 30.4$\pm$1.2 & 18.4$\pm$0.7 & $-$64$\pm$3 & 10$\pm$4 & 3$\pm$4 & $-$64$\pm$39\\
7.95 & 14.33$\pm$0.07& 36.5$\pm$0.8 & 68$\pm$5 & 93 &
 38.1$\pm$2.8 & 18.9$\pm$1.3 & $-$57$\pm$4 & 25$\pm$4 & 4$\pm$6 & $-$51$\pm$14\\
8.28 & 14.02$\pm$0.04& 30.9$\pm$0.5 & 89$\pm$4 & 132 &
 40.0$\pm$1.6 &19.3$\pm$0.8 & $-$53$\pm$2 & 28$\pm$2 & 4$\pm$4 & $-$46$\pm$6\\
8.75 & 14.02$\pm$0.02 & 31.8$\pm$0.3 & 45$\pm$1 & 63 &
 35.6$\pm$0.9 & 20.5$\pm$0.5& $-$50$\pm$2 & 22$\pm$1 & 6$\pm$2 & $-$35$\pm$7\\
9.92 & 14.19$\pm$0.03 & 14.0$\pm$0.4 & 14$\pm$1 & 17 &
 29.9$\pm$1.3 & 20.7$\pm$0.9 & $-$60$\pm$5 & 11$\pm$2 & 7$\pm$5 & $-$5$\pm$67\\
\hline
\end{tabular}
\end{table*}

A direct analysis of the visibility data is another, complementary way
to obtain information about the source structure. Given a model of the
source, this method is able to produce more reliable uncertainties
than fitting a deconvolved image. For a weak source, however, one is
limited to use the simplest source models only as the ambiguity of the
fit increases rapidly with the number of free model parameters.
Fig.~\ref{uvplot} contains the dependencies of the visibility
amplitude on the baseline length for the 7.2~\kss (upper plot) and
$-$1.1~\kss features (lower plot).  Each point in the plot represents
a single baseline. Baselines from the different array configurations
are shown by different symbols.  All visibility data were vector
averaged together separately for each baseline using the positions
shown in Fig.~\ref{orion6specandmap} for the 7.2~\kss and $-$1.1~\kss
features respectively, regardless of the projected length. The
horizontal error bars in Fig.~\ref{uvplot} represent the scatter of
the projected length for each baseline. The vertical error bars show
the scatter of the averaged amplitude. This complicated scheme of
averaging is dictated by the weakness of the source. Different fit
results are shown by the solid and dashed lines in Fig.~\ref{uvplot}.
The 7.2~\kss feature can be described by a symmetric Gaussian source
model with a total flux density of 0.6$\pm$0.1~Jy and FWHM of
18\arcsec$\pm$9\arcsec.  Taking into consideration that the size of
this symmetric model is an average size as the actual source is most
probably asymmetric, the value is in agreement with Table~\ref{fit6}
(deconvolved sizes and separation between peaks at different
velocities). From Fig.~\ref{uvplot} the $-$1.1~\kss feature appears to
have undergone a flare during the EW367 observations (2004 April 10)
as the flux densities are significantly different from other
configurations (including the 6A observations from 2004 January 5).
The solid line represents a point source model with a flux density of
0.5$\pm$0.1~Jy fitted to all data except that from the EW367
configuration. One can see in Fig.~\ref{uvplot} that the $-$1.1~\kss
feature is unresolved for baseline lengths of about
$8\times10^4$~$\lambda$, which implies the source size less than about
2\farcs6. The EW367 data were fitted separately.  A gradual decrease
of the visibility amplitude at the short EW367 baselines seen in
Fig.~\ref{uvplot} encouraged us to try a two component model, although
the point source model is adequate for the data (with the observed
noise level an extended component is not necessary to explain the
plot).  The point source model (short dashed line in
Fig.~\ref{uvplot}) implies the flux density of 1.2$\pm$0.2~Jy. The two
component model (long dashed line) includes an unresolved component
with the flux density of 1.0$\pm$0.2~Jy and a symmetric Gaussian
source with a total flux density of 0.4$\pm$0.1~Jy and the FWHM
17\arcsec$\pm$13\arcsec. Note, that the flux densities and source
sizes in these models are overestimated (less for brighter signals)
due to amplitude bias. However, the magnitude of this systematic
offset is well below the random noise.

\subsection{The 25-GHz data}
\label{the25ghzdata}

The 25-GHz methanol maser in OMC-1 was the first maser from the
methanol molecule detected in space \citep{bar71,chu74}.
\citet{men88b} observed several lines (J$=2$ to 8) in the series and
found that maser emission dominates in the spectra for the J$=4$ line
and higher while most of the emission in the J$=2$ and 3 transitions
is of a thermal (non-maser) nature. Emission was detected from both
Orion-KL and Orion-South, resembling the morphology of the source seen
in mm-wavelength methanol lines \citep{men88a}.  \citet{joh92} imaged
the J$=5$ and 6 masers in the KL region using the VLA D-configuration
and showed that the maser consisted of a large number of spots with a
small velocity spread, distributed in a 40\arcsec{ }long crescent
shaped region.  The spatial resolution of the ATCA in the H75B
configuration is much lower than that of the VLA and is comparable
with the resolution of the single dish Effelsberg map.  Therefore,
unlike \citet{joh92} we cannot disentangle any information about
individual maser spots and here only attempt to analyze a general
picture. From the shape of the spectra on different baselines we have
chosen several spectral channels corresponding to the peaks of
distinct features. Images of the spectral channels at
velocities of 7.34, 7.58, 7.95, 8.28, 8.75, and 9.92~\kss are shown in
Fig.~\ref{orion25maps}.  All these images have been corrected for
primary beam attenuation. The maser spots revealed by \citet{joh92} at
close velocities are overlayed on our 25-GHz image in
Fig.~\ref{orion25maps}.  For each subplot, we show only those spots,
which have peak velocities closer to the subplot velocity than the
half width at half maximum determined by \citet{joh92} for these spots
plus the width of one spectral channel in the \citet{joh92} data
(0.15~\ks).  In Fig.~\ref{orion25maps} we also show the 95-GHz (the
8$_0-7_1$~A$^+$ methanol transition) images sampled at subplot
velocities from the spectral cube obtained by \citet{wri96}, which is
available from the NCSA Astronomy Digital Image Library
(ADIL)\footnote{Accessible via URL \it http://imagelib.ncsa.uiuc.edu}
and has a spectral resolution of 0.5~\ks (our 25-GHz data have a
spectral resolution of 0.05~\ks).  In a similar manner to the 6.7-GHz
data we fitted a Gaussian source model into each spectral channel. The
results are listed in Table~\ref{fit25}.  The columns are the same as
for Table~\ref{fit6}.  As before the uncertainties in the table are
formal errors of the Gaussian fit and a good estimate of the actual
1$\sigma$ accuracy is given by the synthesized beam size and signal to
noise ratio, which gives 2.4\arcsec$\times$1.5\arcsec{ }with a
positional angle of $-$64\degr.

Although we know from the image of \citet{joh92} that the 25-GHz
emission contains a larger number of components, these characteristic
velocities we show are sufficient to analyze the morphology of the
source within the spatial resolution provided by the ATCA.
\citet{men88b} also found 6 components in their Effelsberg data, which
has a comparable spatial resolution to our measurement. However, the
component velocities listed by \citet{men88b} are slightly different.
Moreover, our data imply the presence of a 7.58~\kss feature, which is
not listed by \citet{men88b}, but do not require a 9.6~\kss feature
detected by them. The difference is most probably due to the
interferometric nature of our measurement as the output signal for a
complex source depends on the size of individual baseline and its
orientation with respect to the source. Another plausible explanation
of the difference is temporal variability in the source. The image
of \citet{joh92} does not contain a spot at or near 9.6~\ks.

A comparison of our 25-GHz image with the maser spots of \citet{joh92}
in Fig.~\ref{orion25maps} (see also Table~\ref{fit25}) shows that they
are in general agreement. However, the south-eastern group of maser
spots dominates in our image at low velocities. The most north-western
spot in the 7.34-\kss subplot in Fig.~\ref{orion25maps} is, therefore,
weak according our data. This is the spot number 11 in Table~2 of
\citet{joh92}. This spot is the brightest spot observed (total flux
92~Jy) and is rather broad in the velocity domain (peak is at 8.0~\ks,
half peak linewidth is 1.03~\ks). Therefore, we suspect that
significant temporal variations have occurred in the 25-GHz masers
since the observations of \citet{joh92}.

\begin{figure}
\includegraphics[width=\linewidth]{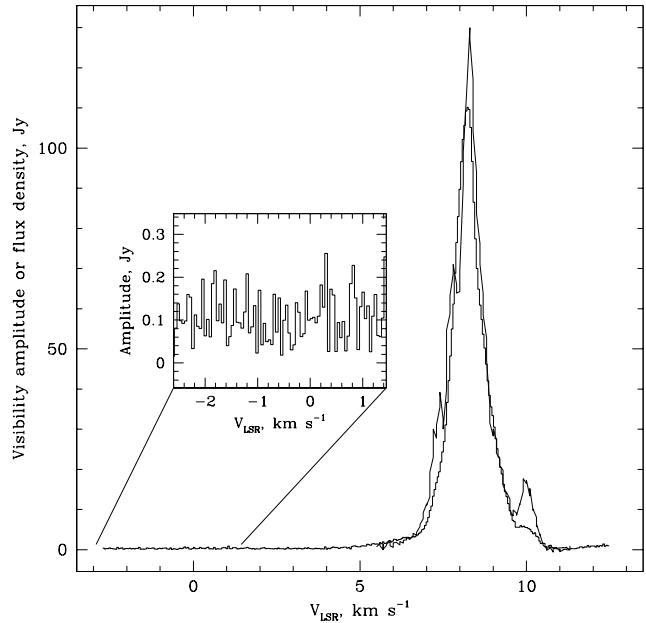}
\caption{The spectrum scaled to correct for the primary beam
attenuation and the spectrum of the image flux density (bold line) of
the 25-GHz maser in OMC-1. The subplot shows the 25-GHz spectrum
at the position of the 6.7-GHz emission near $-$1~\ks. The scaling factors
are 2.71 for the main plot and 1.16 for the subplot.}
\label{orion25spec}
\end{figure}

The vector averaged spectrum of the 25-GHz emission is shown in
Fig.~\ref{orion25spec}.  The data on all baselines except those
including the {\sc CA06} antenna, were vector averaged together for
the position $\alpha_{2000}=5^h35^m$14\fs02,
$\delta_{2000}=-5$\degr22\arcmin30\farcs9, 
which corresponds to the peak of the 25-GHz emission in the image
(Fig.~\ref{orion25maps}) at 8.28~\ks. The amplitude was scaled by a
factor of 2.71 to account for primary beam attenuation at the offset
of the above mentioned position with respect to the pointing centre.
However, this method of correcting for the primary beam is valid only
for a point source. In a complex source the spots, which lie closer to
the pointing centre than the above mentioned position, will appear
brighter, while the effect of the spots which lie further will be
underestimated. Moreover, the measured flux density is very sensitive
to the pointing errors when the source is offset from the pointing
centre.  The magnitude of this effect is determined by the derivative
of the primary beam and results in an error of 17$-$39 per cent (at
the position of the 25-GHz peak) if a pointing error of 5$-$10\arcsec{
}is assumed.  The impact of the pointing errors on the measured flux
density for a complex extended source can be alleviated by proper
mosaicing observations. Another way to obtain a spectrum of the source
from interferometric data is to sum the flux density in the
deconvolved image for each spectral channel. This enables a proper
primary beam correction to be done, taking into account the different
positions of individual features and produces a single dish-like
spectrum where the individual spectral features are emphasized.
However, this method includes a non-linear deconvolution algorithm,
which makes it difficult to analyze the noise and to assess whether a
feature in the spectrum is reliable. In addition, in the case of a
poor uv-coverage the quality of such spectrum is limited by the
dynamic range of the image (5:1 in our 25-GHz measurement) and the
absolute noise level is different at the line centre and at the wings.
Freedom in the choice of the image region where pixel values are
summed and the fact that the values of adjacent pixels usually
correlate make this method inappropriate for weak signals like the
6.7-GHz line discussed in the previous section. However, such a
spectrum can be constructed for the 25-GHz maser which is bright.  The
spectrum obtained by summing over a 100\arcsec{ } square centered at
$\alpha_{2000}=5^h35^m$14\fs33,
$\delta_{2000}=-5$\degr23\arcmin14\farcs6 is shown in
Fig.~\ref{orion25spec} by the bold line. The size and position of the
image region involved in summation is chosen to accomodate all the
emission seen in Fig.\ref{orion25maps} and to avoid as far as possible
the region near the edge of the field of view where the noise of
deconvolution algorithm is amplified by the primary beam correction
procedure. The most significant difference between this spectrum and
the vector averaged one is near 10~\ks. This is because the source is
offset to the north at these velocities (see Fig.~\ref{orion25maps})
and the primary beam correction uses the position of the peak at
8.28~\kss~(see the discussion above).  The spectrum resembles that of
\citet{joh92} obtained using the same method, although our spectrum
peaks at 8.28~\kss, while that of \citet{joh92} peaks at about
7.9~\ks, and an additional feature near 7.6~\kss seems to be present
in our spectrum.  These differences can be explained by both the
errors resulting from the dynamic range and the pointing inaccuracy as
well as by different uv-coverage in the two experiments.  Temporal
variability of the maser is another possible explanation.  The peak
flux density determined using both the vector averaged spectrum and
the image flux spectrum is in agreement with observations of
\citet{joh92} and \citet{men88b} within the accuracy of our
measurement.

\citet{men88b} report 25-GHz methanol emission from the Orion-South
region (southern 6.7-GHz source) for the J$=2,3$ and 4 lines in the
series. The source has also mm-wavelength methanol line emission
peaking at 6~\kss\citep{men88b}.  However, no methanol emission at
$-$1.1~\kss has been reported. The subplot in Fig.~\ref{orion25spec}
shows the 25-GHz vector averaged spectrum at velocities near
$-$1.1~\ks. As for the main plot, all data except those resulting from
the baselines with the {\sc CA06} antenna were vector averaged
together. However, for the subplot in Fig.~\ref{orion25spec} the
position of the averaging centre was
$\alpha_{2000}=5^h35^m$13\fs93, $\delta_{2000}=-5$\degr24\arcmin05\farcs3, 
which correspond to the position determined for the 6.7-GHz emission
at $-$1.1~\ks (Fig.~\ref{orion6specandmap}). The pointing centre was
much closer to this position and, therefore, a smaller primary beam
correction factor of 1.16 has been applied. The flux density of the
$5_2-5_1$~E emission at 25~GHz near the velocities of $-$1.1~\kss is
below the detection limit. The rms noise (1$\sigma$) is 0.12~Jy.

\section{Discussion}
\subsection{Temporal variability and previous observations}
The non-imaging analysis discussed in~\ref{result6} shows that the
$-$1.1~\kss feature was stronger in the EW367 data (2004 April 10)
with respect to the data from other array configurations. This
difference is unlikely to be due to instrumental effects. The derived
secondary calibrator flux densities at 6.7-GHz are 0.68~Jy, 0.87~Jy,
0.87~Jy, and 1.07~Jy for the 2003 June 9$-$10 (H75B), 2003 November 12
(1.5D), 2004 January 5 (6A), and 2004 April 10 (EW367) observations,
respectively.  These numbers imply that the precision of the absolute
calibration is better than 16 per cent. Although the secondary calibrator
appears brighter in the EW367 data, the difference in the flux density
is not enough to explain the $-$1.1~\kss feature being 2$-$3 times
stronger in these data.  In addition, there is a good agreement
between the amplitudes on short and long baselines of the same session
(Fig.~\ref{uvplot}). Hence, a difference in the phase stability at
different dates is unlikely to explain the observed difference of
amplitudes (a vector averaging of the data is used before the
amplitudes are calculated) because the decorrelation should be
stronger on longer baselines. A final argument against instrumental
effects is an agreement between the 7.2~\kss feature data from the
different array configurations. Therefore, it is likely that the
$-$1.1~\kss feature had a flare in April 2004.

The first observations of the OMC-1 at 6.7~GHz were carried out by
\citet{cas95} on 1992 December 17$-$24 using the Parkes 64-m
telescope.  The telescope was pointed towards
$\alpha_{2000}=5^h35^m$14\fs35,
$\delta_{2000}=-5$\degr22\arcmin24\farcs4. A weak (0.5~Jy) feature at
7~\kss was reported. The detection limit was about 0.3~Jy \citep{cas95}.
So the flux density measured more than a decade ago at Parkes for the
7.2~\kss feature is in good agreement with our ATCA observations (see
Table~\ref{fit6}). The Parkes spectrum (Caswell, priv. comm.) shows
an absorption feature at 4$-$7~\ks, which is not present in our data.
This absorption obscures the red wing of the emission feature
and is probably resolved by the ATCA. The Parkes spectrum
also contains a hint on the $-$1.1~\kss feature at the level of 0.2~Jy.
According to our measurement (Table~\ref{fit6})
the $-$1.1~\kss feature is located approximately 1\farcm6 to the south
from the position observed by \citet{cas95}. Such an offset implies a
beam attenuation factor of 2, assuming the half-power beamwidth of
3\farcm3 at 6.7~GHz.  If the flux density of the $-$1.1~\kss
feature was 0.5~Jy as we had in all sessions except April 10
(Fig.\ref{uvplot}), the attenuated flux density would be in agreement
with the observations of \citet{cas95}.

The C425 project from the ATCA archive was devoted to search for
methanol masers at 6.7~GHz in proplyds in the Orion nebula. The
observations were carried out on 1995 January 6 and March 30 using the
6A and 1.5A array configurations respectively. The phase and pointing
centre was at $\alpha_{2000}=5^h35^m0^s$,
$\delta_{2000}=-5$\degr23\arcmin0\arcsec, which is about 3\farcm7 away
from the positions of the 6.7-GHz emission we have measured
(Table~\ref{fit6}). The offset of 3\farcm7 implies a factor of 2
attenuation due to the primary beam. We processed these data the same
way as our 6.7-GHz observations and created spectra similar to
Fig.~\ref{orion6specandmap}. The amplitudes were scaled to correct for
the primary beam attenuation. Neither the $-$1.1~\kss feature, nor the
7.2~\kss were detected in the January 6 data which had a typical
1$\sigma$ single baseline amplitude rms of about 0.2~Jy. From the
calibrator data we suspect serious technical problems during this
session.  The March 30 data suffer from poor phase stability, however,
both spectral features are clearly present in the data. The 7.2~\kss
feature has an amplitude of about 0.6$\pm$0.2~Jy on the shortest
baselines, which is in agreement with our measurement
(Table~\ref{fit6}, Fig.\ref{orion6specandmap} and~\ref{uvplot}).  The
$-$1.1~\kss feature has an amplitude of about 0.2$\pm$0.1~Jy and is
revealed only if all baselines are averaged together. This value is
much lower than the flux density of about 0.5~Jy expected from our
observations, although it is underestimated because baselines of
different lengths have been averaged together and the phase stability
was bad.  However, we cannot exclude the possibility that the
$-$1.1~\kss feature was really fainter in 1995, than in 2003, as we
have observed it to be varible on a timescale of months.

\subsection{The 6.7-GHz emission at 7$-$8~\ks}

Inspection of Tables~\ref{fit6} and \ref{fit25} as well as
Figure~\ref{orion25maps} show that the 6.7-GHz emission at 7$-$8~\kss
is coincident with the 25-GHz emission detected in our observations
within the uncertainty of the measured positions. The tendency to move
to the east for lower velocity exists for the transitions at both
frequencies.  In Fig.~\ref{orion25maps} we overlayed the 25-GHz spots
found by \citet{joh92}, which have more accurate positions than our
measurement but were observed 15 years earlier. Also superimposed are
the 95-GHz image obtained by \citet{wri96} and 3$\sigma$ error
ellipses showing the positions of the 6.7-GHz emission at 7.2~\ks, and
7.5~\ks.  It is clear from Fig.~\ref{orion25maps} that both components
of the 7$-$8~\kss feature may be associated with 25-GHz spots,
although the agreement is better for the 7.2-\kss component (left
ellipse in Fig~\ref{orion25maps}) in terms of both position and
velocity. This component is likely to be associated with the
south-eastern tip of the crescent shaped region traced by the 25-GHz
maser spots.  The uncertainty ellipse for the 7.5-\kss component may
include the southern peak at 95-GHz. Although the coarse spectral
resolution in the data of \citet{wri96} makes this comparison
difficult.  The 7$-$8~\kss feature consists of at least two components
emanating from close locations. In this case the Gaussian fit may
suggest the positions of these components to be closer together than
they are in reality. Therefore, the actual agreement may even be
better.  The source is too complex to determine any definite
association with the spatial resolution of our 6.7-GHz data, but the
general coincidence with the area containing 25-GHz and 95-GHz
emission is obvious.  A high resolution study is required to measure
the position of the 6.7-GHz emission with respect to individual 25-GHz
maser spots. However, it is difficult to do this with existing
interferometers because the source is very weak at 6.7~GHz and most
probably resolved.

It is clear from Fig.~\ref{orion25maps} that the coincindence between
25-GHz maser spots revealed by \citet{joh92} and the peaks of the
95-GHz emission detected by \citet{wri96} is good, but not perfect. It
seems unlikely that this small discrepancy is due to positional
errors.  Most probably, the 25-GHz masers require more specific
physical conditions than the 95-GHz masers. The bright masers at
25-GHz are relatively rare \citep{men86b}, while 95-GHz masers are
common \citep[e.g.,][]{val00}. Another possible explanation is
temporal variability of the 25-GHz emission suspected in section
\ref{the25ghzdata}.  It is worth noting, that the nature of the 95-GHz
emission in OMC-1 is not completely clear. This is a well known
class~I maser transition \citep[e.g.,][]{val00} and \citet{pla88}
concluded on the basis of the brightness temperature that at least a
part of the 95-GHz emission in OMC-1 is due to a maser action.
However, the limit on the brightness temperature ($T_b>500$~K) set by
\citet{pla88} is not strong and higher spatial resolution observations
are required to answer the question unambiguously.  A comparison of
interferometric images of the 25-GHz and 95-GHz masers with that of
various outflow and dense molecular gas tracers shown that these
masers arise in the interface regions where outflows interact with the
ambient material \citep{joh92,pla90}. Therefore, the 7$-$8~\kss
feature is likely to emanate from the same volume of gas disturbed by
the outflow, even if there is no co-existense with the 25-GHz maser
spots at finer scales.

At present we cannot definitively discriminate whether the
observed 6.7-GHz line peaking at 7.2~\kss is a maser or has a
quasi-thermal nature (i.e., the level populations of the 6.7-GHz
transition are without inversion). The straightforward approach to
estimate the brightness temperature limit using the fit result at
either velocity listed in Table~\ref{fit6} gives a brightness
temperature of about 40~K. Because the feature definitely is
resolved by our observations, this value is a rough estimate of
the average brightness temperature and should always be considered
as a lower limit to the actual brightness temperature.

\subsection{Co-existence of masers belonging to different classes}
\label{pumping}

In the previous section we have shown that the 6.7-GHz (class~II maser
transition) emission feature peaking at 7.2~\kss is likely to arise
from the region where the 25~GHz masers are (class~I). The spatial
resolution of our data is too coarse to be able to claim that exactly
this is realized in such a complex source as OMC-1, although there is
agreement within the positional uncertainty of our measurement. This
is a new finding because, as mentioned above, methanol masers of
different classes co-existing in the same star-forming region are
usually situated at considerably different locations in the source.

In this section we address the possibility of coincidence of the
different class maser locations in the strict theoretical sense:
whether there is a pumping regime allowing a simultaneous inversion of
transitions belonging to different classes in the same volume of gas?
Previous studies show that this is possible in principle. For example
the class~I $11_2-11_1$~E transition at 26.3~GHz is present in the
list of class~II methanol maser candidates of \citet{sob97b}. However,
the existence of inversion in class II maser transitions under
conditions where class I masers are strong has not been seriously
considered, though it was briefly mentioned in the conference
proceedings by \citet{vor05}.

Pumping analysis has shown that the dust temperature of the maser
environment is one of the key parameters which determines the
relative strength of the different types of methanol maser
transitions. When the dust temperature is high ($T>120$~K) the
radiative excitation from the ground to the second torsional state
becomes influential and controls the populations of the masing
transitions. The levels of the first and the second torsionally excited
states are located $350-650$~K and $600-800$~K above
the levels of the ground state, respectively \citep{mek99}.
Therefore, the corresponding transitions
occur at $(0.6-1.3)\times10^{13}$~Hz ($20-40~\mu$m) and 
$(1.2-1.7)\times10^{13}$~Hz ($15-25~\mu$m).
The strong class II masers
appear when the dust temperature is sufficiently high ($T>150$~K)
and the gas temperature is lower or about the same
\citep{sob94,sob97a}. Calculations have shown that the strong
class I masers do not appear when the distribution of population
numbers is controlled by radiative processes in transitions
between the ground and the second torsionally excited state. The
situation changes only for rather high densities (exceeding
$10^8$~cm$^{-3}$) which are not likely to occur in methanol maser
sites.

The situation at lower temperatures was considered by \citet{cra92}.
This work was limited to the case where conditions in the masing
region are such that the radiative transitions between different
torsionally excited states are not influential.  This implies dust
temperatures below 50 K. \citet{cra92} have shown that the class~I
maser transitions are considerably inverted only when the gas
temperature exceeds that of the ambient radiation, i.e., excitation
processes are dominated by collisions with the gas. The class~II
masers appear only when the radiative excitation prevails.  So
situations where the 6.7-GHz transition can attain considerable
inversion in the star forming region simultaneously with the strong
class I masers, are limited to cases where the dust temperature is in
the range between 50~K to 120~K and the gas temperature has a similar
value.

\begin{figure}
\includegraphics[width=\linewidth]{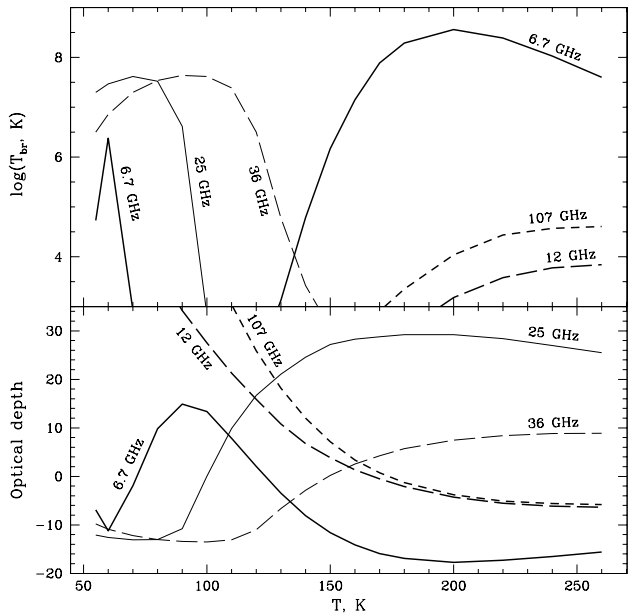}
\caption{The model dependence of the brightness temperature and
optical depth of various methanol
maser transitions on the temperature ($T=T_{gas}=T_{dust}$).
Bold lines represent class~II transitions.}
\label{model}
\end{figure}

We considered a set of models where the dust responsible for the
pumping is intermixed with the gas within the molecular cloud. The
dust temperature was varied over the range 50~K to 260~K.  Figure 5
displays the results of calculations for models with different dust
(T$_{dust}$) and gas (T$_{gas}$) temperatures, which were assumed to
be equal. This is a good approximation for regions with density and
temperature in the range typical for the 25-GHz masers \citep[see,
e.g.,][]{sob03}, which appear after the passage of a moderate shock
\citep[e.g.,][]{whi97}. Outlines of the code are described in
\citet{sut04} and further details can be found in
\citet{sob97a,sob97b} and \citet{koe80}. The main difference from the
modelling procedure described in \citet{sob97b} is that the new
calculations take into account radiative processes on the dust which
is intermixed with the gas within the source, and the external
radiation is reduced to the microwave background with $T_{bg}=2.7$~K.
The cloud models presented have the following parameters: the hydrogen
number density is $10^5$~cm$^{-3}$, the specific (per unit of the
linewidth) methanol column density is $10^{12}$ cm$^{-3}$s, and the
beaming factor ($\varepsilon^{-1}$) is 10. For all models shown here
the optical depth of the dust in the source is set to be proportional
to the square of the frequency and assumed to be equal to 0.5 at
$10^{13}$~Hz (30$\mu$m). The dependence of the
brightness temperature and optical
depth in different transitions versus the temperature
($T=T_{gas}=T_{dust}$) is shown in Fig.\ref{model}. A maser line is
expected to have a high brightness temperature and a negative optical
depth.  Increasing the temperature ($T$) in such a cloud changes the
balance of excitation processes in favour of radiative ones.  So, the
expectation is that the class~I masers should appear at low
temperatures while at high temperatures the class~II masers should
shine brightly. The transitions all show a single smooth change from
non-masing to masing condition (positive to negative optical depth),
with the notable exception of the 6.7-GHz transition. The secondary
masing regime of the 6.7-GHz transition is at low temperature (60~K),
overlapping with the regime favouring the class~I 25-GHz masers.  This
is a new pumping regime, which is not expected in the traditional
methanol maser classification scheme.

To investigate this regime further, we repeated calculations for a
single temperature $T=T_{dust}=T_{gas}=60$~K using a modified energy
level diagram. We deliberately excluded the energy levels of the
torsionally excited states of the methanol molecule to check their
influence on the overall pumping.  In addition, in some models we
simulated the pumping by an external dust layer as it was done by
\citet{sob97b}. For these models an isotropic distribution of the dust
emission was assumed (i.e. the dilution factor is 1).  All other
parameters were the same as in the models used to produce
Fig.~\ref{model}. More detailed analysis require a comprehensive
investigation of the vast parameter space and will be reported
elsewhere. The optical depths of selected transitions obtained in
these models are listed in Table~\ref{modtab}.  The first model
corresponds to the $T=60$~K point in Fig.~\ref{model}.
Table~\ref{modtab} shows that the 6.7-GHz transition is inverted only
if the first torsionally excited state is included. Addition of the
second torsionally excited state does not siginificantly affect the
result, except for model~4 involving an external dust layer.  This
means that the pumping of the 6.7-GHz transition in this low
temperature regime is going through the levels of the first
torsionally excited state in contrast to the pumping of the bright
class~II masers, which requires both the first and second torsionally
excited states \citep{sob94}.  In the models where the dust is
intermixed with gas (labelled I in Table~\ref{modtab}), the 6.7-GHz
transition is inverted stronger (greater optical depth by absolute
value) than in the case of the external dust (labelled E in
Table~\ref{modtab}). In model~4, which includes both the first and
the second torsionally excited states and simulates the pumping by an
external dust layer, the 6.7-GHz transition is not inverted.
Therefore, it is essential that the dust is intermixed with the gas
for this low temperature mechanism to be realized in practice.  There
is no significant difference between the optical depths in the class~I
maser transitions at 25~GHz and 36~GHz in these models. This means
that these transitions are pumped mostly through the levels of the
ground torsional state of the molecule. For the given values of the
fixed model parameters, the transitions at 44~GHz and 95~GHz are not
inverted if at least one torsionally excited state is included.  These
transitions belong to the $(\mathrm{J}+1)_0-\mathrm{J}_1$~A$^+$
transition series (J=6 and 7, respectively), which occurs between the
same K-ladders as the $5_1-6_0$~A$^+$ transition at 6.7~GHz, but in
the opposite directions.  Therefore, the tendency for 6.7~GHz to
invert due to the pumping through the levels of the first torsionally
excited state reduces the inversion of the
$(\mathrm{J}+1)_0-\mathrm{J}_1$~A$^+$ transitions.  Finding the right
balance between these processes requires a detailed study of the
parameter space.

\begin{table}
\caption{Optical depths of selected transitions for various models.
For all models T=60~K. The first 3 models involve the dust intermixed with
the gas in the cloud (I), the last 3 models involve an external
isotropic layer of dust~(E).}
\label{modtab}
\begin{tabular}{crrrrrr}
\hline
\multicolumn{1}{r}{Model}&\cthead{1}&\cthead{2}&\cthead{3}&
\cthead{4}&\cthead{5}&\cthead{6}\\
\multicolumn{1}{r}{Dust}&\cthead{I}&\cthead{I}&\cthead{I}&
\cthead{E}&\cthead{E}&\cthead{E}\\
\multicolumn{1}{r}{$v_t=1$}&\cthead{yes}&\cthead{yes}&\cthead{no}&
\cthead{yes}&\cthead{yes}&\cthead{no}\\
\multicolumn{1}{r}{$v_t=2$}&\cthead{yes}&\cthead{no}&\cthead{no}&
\cthead{yes}&\cthead{no}&\cthead{no}\\
\hline
\cthead{Frequency}&\multicolumn{6}{c}{Optical depths}\\
\cthead{GHz}&\multicolumn{6}{c}{}\\
\hline
 6.7 & $-$15.7 & $-$16.6 & 142 & 18.1 & $-$11.5 & 139\\
 12  & 68.3    & 68.9    & 192 & 100 & 116     & 195\\
 25  & $-$12.7 & $-$12.9 & $-$10.8 & $-$11.6 & $-$11.8 & $-$10.8\\
 36  & $-$11.3 & $-$11.3 & $-$10.5 & $-$10.4 & $-$10.7 & $-$10.5\\
 44  & 77.1    & 89.6    & $-$8.9  & 21.9 & 49.9    & $-$8.9 \\
 95  & 101     & 113     & $-$1.0  & 49.8 & 72.0    & $-$1.0 \\
 107 & 48.0    & 33.6    & 192     & 49.8 & 61.5    & 190 \\
\hline
\end{tabular}
\end{table}

\subsection{The $-$1.1~\kss feature at 6.7~GHz}

According to position listed in the Table~\ref{fit6}, the $-$1.1~\kss
feature at 6.7~GHz originates in the Orion-South region, which contains a
prominent dust and NH$_3$ condensation \citep{kee82,bat83}. However,
the feature is noticeably blue-shifted with respect to the normal
spread of velocities for lines tracing this molecular core and no
emission in methanol lines has been previously reported in Orion-South at
this velocity \citep[see, e.g.,][]{tat93,men88a}.  The source appears
to be unresolved (Table~\ref{fit6}) which implies an angular size less
than 5\arcsec$\times$2\arcsec{ } and a brightness temperature
greater than about 1400~K. The non-imaging analysis of the visibility
data described in~\ref{result6} yields the size less than 2\farcs6 (a
fit for all data except that from the EW367 array configuration), and
thus the brightness temperature greater than about 2000~K. This high
brightness temperature and peculiar velocity of the feature indicate
that the detected line most probably is a maser. The feature appears
to show temporal variability (see section~\ref{result6}), which is
another indicator that it is likely a maser and not quasi-thermal.

Sensitive observations in the infrared and radiocontinuum
\citep{zap04,ode03} do not show the presence of a compact source at
the peak position of the newly detected methanol maser listed in the
Table~\ref{fit6}.  However, positional uncertainties of our
observations do not allow us to rule out the possibility of
association with 1.3 cm sources 139-409 and 140-410 of \citet{zap04}.
The nature of these sources is unclear and they may be sources of
gyrosynchrotron emission or ionized outflows from low-mass stars
\citep{zap04}.

It is likely that the methanol maser in Orion-South is associated with
the outflow.  \citet{sch90} detected a CO outflow presumably ejected
from the CS3/FIR4 infrared source. The CS3 and FIR4 positions may
actually represent the same object observed in the CS line and 1.3~mm
continuum, respectively.  The nominal separation between them is
5\farcs5 along the outflow axis \citep{sch90}. The receding part of
the outflow is a remarkable 120\arcsec{ } long jet with a very narrow
width (8\arcsec).  The approaching part of the outflow appears on the
other side of CS3/FIR4 and has an irregular shape covering the whole
Orion-South region. The component is seen at velocities from 0 to
3~\kss and encompasses the methanol maser location.  \citet{sch90}
argued that this component of the outflow might protrude into the {\sc
  HII} region M42, which could explain its ill-defined appearance.
Observations of SiO lines tracing dense parts of the outflow display
blue wings spanning down to $-$5~\kss\citep{ziu90}. The position of
the methanol maser at $-$1.1~\kss coincides with the blueshifted SiO
emission within the uncertainty of the measurement
(Table~\protect\ref{fit6} and Ziurys et al. \citeyear{ziu90}).  It is
therefore possible, that the maser is associated with the approaching
part of the outflow from CS3/FIR4.

It should be mentioned that the Orion-South region contains several
outflows \citep[see, e.g.,][]{ode03} possibly originating from the
young stellar objects whose formation was induced by outflow described
in \citet{sch90}.  Observations of H$_2$O masers \citep{gau98} and HH
objects \citep{ode03} show that outflows in the region are quite rapid
and the velocity of the newly detected 6.7-GHz methanol maser line
component is well within their velocity spread. At present it is not
possible to associate this methanol maser with any particular outflow
or young stellar object.  Another weak 6.7-GHz methanol maser probably
associated with an outflow has been found towards NGC~2024 (source
FIR4) by \citet{min03}.  Though NGC~2024 is situated rather far from
Orion-South (about 4\degr{ }to the north) it also belongs to the Orion
complex.

The excitation analysis of \citet{sob97a} shows that class~II
methanol masers can be situated at considerable distances from the
young stellar object if the associated region is affected by an
outflow which releases methanol from the icy dust grain mantles
and warms up the dust in the maser vicinity. Great enhancement of
the methanol abundance at the edges of the outflows is confirmed
observationally \citep[see, e.g.,][]{bac95,bac98,gib98}.
Association of a considerable portion of class~II methanol masers
with outflows is supported by statistics \citep[see,
e.g.,][]{mal03,cod04} and observations of particular sources
\citep[see, e.g.,][]{deb02,min03}. We are currently unable to
reveal whether these weak masers found in the Orion complex are
pumped via a low temperature mechanism similar to that described
in \ref{pumping} or by the standard scheme
\citep{sob94,sob97a,sob97b} using the hot dust emission from a
nearby source. According to the position listed in
Table~\ref{fit6}, the $-$1.1~\kss maser is associated with the
southern tip of the dust condensation in Orion-South (Mezger, Wink
\& Zylka \citeyear{mez90}). The bulk of the dust in this region is
rather cold (T$<70$~K) to allow the pumping via the standard
scheme \citep{mez90,moo00}. However, condensations with
significantly enhanced dust temperatures ($\approx$200~K) do exist
\citep[e.g.,][]{mez90}. Due to the lack of the dust temperature
maps at high spatial resolution, it is impossible to discriminate
between the pumping regimes on the basis of the dust temperature.
We detected no 25-GHz emission towards the position of the 6.7-GHz
maser in Orion-South. Therefore, if the low temperature mechanism
is realized, the parameters are unlikely the same as used in
\ref{pumping}. Sensitive observations of the other methanol and
hydroxyl maser transitions toward 6.7-GHz maser positions in
Orion-South and NGC~2024 may elucidate physical status of these
sources.

\section{Conclusions}
\begin{enumerate}
\item Two 6.7-GHz methanol emission features have been detected
  towards OMC-1, one narrow feature with a peak at $-$1.1~\kss and a
  broad feature with a peak at 7.2~\ks.
\item The 7.2~\kss feature consists of at least 2 components which are
  close in both position and velocity. These components are likely to
  be associated with the outflow traced by the 25-GHz masers and
  95-GHz emission.
\item There is a pumping regime which produces a weak 6.7-GHz maser
  (class~II) simultaneously with the 25-GHz masers (class~I) for relatively
  low dust temperatures ($T\approx 60$~K). The pumping occurs through the
  levels of the first torsionally excited state and is due to the emission
  of dust intermixed with gas. Brightness temperatures
  of the order of $10^6$~K can be achieved for such masers.
\item The $-$1.1~\kss feature is most probably a maser with a
  brightness temperature $>$1400~K. It originates in the Orion-South
  region and may be associated with the approaching part of the
  outflow seen in CO. This feature is most probably variable.
\item No $5_2-5_1$~E 25-GHz emission has been detected towards the
  Orion-South region at velocities near $-$1~\ks.
\end{enumerate}

\section*{Acknowledgments}
We would like to thank the local staff of the Narrabri Observatory,
and particularly Robin Wark, Bob Sault and Mark Wieringa, for the help
during observations shortly after the new 12-mm system has become
available at the ATCA and for numerous useful advises during the data
processing stage. We greatly appreciate the efforts of James Caswell for
the internal refereeing of the first version of manuscript.
The Australia Telescope is funded by the
Commonwealth of Australia for operation as a National Facility managed
by CSIRO.  The authors would like to thank the NCSA Astronomy Digital Image
Library (ADIL) for providing 95-GHz images obtained by \citet{wri96}
for this article.
MAV was partially supported by the RFBR grants no.
98-02-16916 and no. 01-02-16902 and by the program "Extended objects
in the Universe-2003". AMS and ABO were supported by the RFBR grant
no. 03-02-16433. SPE acknowledges financial support for this work from
Australia Research Council.

\bsp

\label{lastpage}

\end{document}